\documentclass{article}

\usepackage{arxiv}

\usepackage[utf8]{inputenc} 
\usepackage[T1]{fontenc}    
\usepackage{hyperref}       
\usepackage{url}            
\usepackage{booktabs}       
\usepackage{amsfonts}       
\usepackage{amsmath}        
\usepackage{nicefrac}       
\usepackage{microtype}      
\usepackage{lipsum}
\usepackage{graphicx}
\graphicspath{ {./images/} }

\title{Precise Liver Tumor Segmentation in CT Using a Hybrid Deep Learning--Radiomics Framework}

\author{
 Xuecheng Li\\
  School of Information Science \& Engineering, Shandong Normal University\\
  Jinan 250358, China \\
   \And
 Weikuan Jia \\
  School of Information Science \& Engineering, Shandong Normal University\\
  Jinan 250358, China \\
  \And
  Komildzhon Sharipov\\
  Tajik State University of Law, Business and Politics\\
  Sughd 735700, Tajikistan\\
  \And
  Alimov Ruslan\\
Tajik State University of Law, Business and Politics\\
  Sughd 735700, Tajikistan\\
  \And
  Lutfuloev Mazbutdzhon\\
  Tajik State University of Law, Business and Politics\\
  Sughd 735700, Tajikistan\\
  \And
  Ismoilov Shuhratjon\\
    Tajik State University of Law, Business and Politics\\
  Sughd 735700, Tajikistan\\
  \And
 Yuanjie Zheng \\
  School of Information Science \& Engineering, Shandong Normal University\\
  Jinan 250358, China \\
}

\begin{document}
\maketitle
\begin{abstract}
Accurate three-dimensional delineation of liver tumors on contrast-enhanced CT is a prerequisite for treatment planning, navigation and response assessment, yet manual contouring is slow, observer-dependent and difficult to standardise across centres. Automatic segmentation is complicated by low lesion–parenchyma contrast, blurred or incomplete boundaries, heterogeneous enhancement patterns, and confounding structures such as vessels and adjacent organs. We propose a hybrid framework that couples an attention-enhanced cascaded U-Net with handcrafted radiomics and voxel-wise 3D CNN refinement for joint liver and liver-tumor segmentation. First, a 2.5D two-stage network with a densely connected encoder, sub-pixel convolution decoders and multi-scale attention gates produces initial liver and tumor probability maps from short stacks of axial slices. Inter-slice temporal consistency is then enforced by a simple three-slice refinement rule along the cranio–caudal direction, which restores thin and tiny lesions while suppressing isolated noise. Next, 728 radiomic descriptors spanning intensity, texture, shape, boundary and wavelet feature groups are extracted from candidate lesions and reduced to 20 stable, highly informative features via multi-strategy feature selection; a random forest classifier uses these features to reject false-positive regions. Finally, a compact 3D patch-based CNN derived from AlexNet operates in a narrow band around the tumor boundary to perform voxel-level relabelling and contour smoothing. On a multi-centre in-house cohort of 300 liver cancer CT studies, the proposed method achieves an average sensitivity of $0.89 \pm 0.02$, positive predictive value of $0.92 \pm 0.02$, and Dice coefficient of $0.88 \pm 0.03$, outperforming the second-best competing method by 0.03 in sensitivity, 0.02 in PPV and 0.04 in Dice. Comparable results are obtained on the public LiTS dataset, indicating good robustness and cross-dataset generalisation. The overall design is modular and generic, and can be readily adapted to the segmentation of other abdominal organs and tumors.
\end{abstract}


\section{Introduction}

Primary liver cancer remains a major global health burden and is consistently ranked among the leading causes of cancer-related mortality. A large proportion of patients are diagnosed at an intermediate or advanced stage, at which point treatment decisions critically depend on an accurate understanding of the number, size, and spatial distribution of intrahepatic lesions. Cross-sectional imaging, in particular contrast-enhanced computed tomography (CT), is therefore deeply embedded in current clinical workflows for diagnosis, staging, treatment planning, and follow-up. In this context, reliable three-dimensional delineation of both the liver parenchyma and individual tumors is indispensable for tasks such as volumetric assessment, determining resection margins, planning ablative therapies, and evaluating treatment response.

In routine practice, liver and tumor boundaries are usually traced manually or semi-automatically by experienced radiologists on a slice-by-slice basis. Although experts can achieve high-quality contours, this process is time-consuming, prone to inter- and intra-observer variability, and difficult to standardize across institutions. Moreover, the imaging characteristics of liver tumors pose intrinsic challenges to segmentation algorithms. First, contrast between lesions and surrounding parenchyma can be very low, especially for hypovascular or infiltrative lesions in the portal-venous phase. Second, tumor boundaries are often irregular and partially obscured by partial-volume effects, motion artefacts, or adjacent high-intensity structures such as vessels and bile ducts. Third, lesion size ranges from sub-centimetre nodules to bulky masses, and enhancement patterns vary widely across tumor types and treatment stages. Finally, CT data acquired from different scanners and centres exhibit substantial heterogeneity in voxel spacing, reconstruction kernels, noise characteristics, and contrast protocols, all of which may degrade the robustness of purely data-driven models.

Over the past decade, a wide spectrum of approaches has been explored for liver and liver-tumor segmentation, ranging from classical deformable models and graph-based methods to machine-learning approaches based on handcrafted features, and more recently to fully convolutional neural networks (FCNs) and their variants. While modern deep architectures such as U-Net and H-DenseUNet have significantly improved the state of the art by learning rich hierarchical representations from large training sets, they still face several practical limitations in this application. First, standard encoder–decoder networks tend to bias towards large structures and may miss tiny, low-contrast lesions unless carefully regularized or augmented. Second, single-stage FCN models typically treat all voxels within the field of view equally and may produce unnecessary false positives along organ boundaries or in regions of imaging noise. Third, purely deep-learning-based methods often operate as “black boxes” and may not fully exploit complementary domain knowledge captured by handcrafted radiomic descriptors, which are designed to characterize lesion shape, intensity distribution, and heterogeneity in a model-agnostic fashion.

These observations motivate a hybrid strategy that combines the strong representation power of deep convolutional networks with the interpretability and robustness of radiomics, as well as explicit voxel-level refinement in three dimensions. In this work, we propose a liver and liver-tumor segmentation framework tailored to portal-venous phase CT that follows this philosophy. At its core lies a 2.5D cascaded U-Net with dense connectivity and attention gates, which jointly segments the liver and tumors while leveraging anatomical priors from the liver branch. To further reduce false positives and improve contour quality, we introduce a radiomics-based lesion verification module and a compact 3D CNN that operates in a narrow band around the current tumor boundary. The resulting system explicitly separates three complementary tasks—coarse localization, candidate verification, and fine-grained boundary adjustment—yet remains computationally efficient and practical for deployment.

The main contributions of this paper are summarised as follows:

\begin{itemize}
  \item \textbf{Cascaded 2.5D attention-dense U-Net.} We design a cascaded segmentation network in which a liver-focused branch and a tumor-focused branch share a densely connected encoder backbone while operating on short stacks of neighbouring slices. The decoder employs sub-pixel convolution for resolution recovery and a set of multi-scale attention gates that selectively amplify features in likely lesion regions. This combination enhances multi-scale context aggregation and substantially improves sensitivity for small and low-contrast tumors.

  \item \textbf{Temporal consistency-aware post-processing.} Beyond standard morphological cleaning, we introduce a lightweight refinement scheme that enforces three-slice consistency along the cranio–caudal direction. By leveraging the strong anatomical continuity of abdominal CT volumes, the method restores thin or tiny tumor fragments that might be removed by erosion while suppressing isolated noisy voxels, thereby improving stability of the final segmentation.

  \item \textbf{Radiomics-driven false-positive suppression.} For each candidate lesion region produced by the cascaded network, we compute a comprehensive set of handcrafted radiomic features covering intensity statistics, texture, shape, boundary characteristics, and multi-scale wavelet descriptors. A multi-strategy feature selection procedure followed by a random forest classifier is used to distinguish true lesions from false positives. This explicit verification stage reduces spurious detections around organ interfaces and vessels without sacrificing sensitivity.

  \item \textbf{Voxel-level boundary refinement using a 3D CNN.} To further sharpen tumor contours, we employ a compact 3D CNN, inspired by AlexNet, trained to classify voxels in a narrow band surrounding the current boundary based on local 3D patches. This module focuses computational effort on the most ambiguous regions and yields smoother, more anatomically plausible contours while keeping the overall model size and inference time moderate.

  \item \textbf{Comprehensive evaluation and practical deployment.} The full pipeline is evaluated on a multi-centre in-house dataset of 300 liver cancer CT cases and on the public LiTS benchmark, demonstrating consistent improvements over several strong baselines in terms of sensitivity, positive predictive value (PPV), and Dice coefficient. The method has been integrated into a prototype clinical research platform supporting 3D visualization and quantitative analysis, illustrating its potential for translation into real-world workflows.
\end{itemize}

By explicitly combining deep learning, radiomics, and 3D voxel-level refinement in a unified framework, the proposed approach aims to bridge the gap between high segmentation accuracy on challenging CT data and the robustness and interpretability required for clinical adoption.

\section{Related Works}
Liver tumor segmentation in CT images can be viewed as a special case of medical image segmentation and has been tackled using a variety of traditional and learning-based methods.

\subsection{Traditional and Classical Machine Learning Methods}
Early studies primarily relied on classical image processing and deformable models. Chlebus et al.~(2005)\cite{lu2005liver} employed an active contour model for semi-automatic liver tumor volume estimation. Manjunath \& Kwadiki et al.~(2008)\cite{choudhary2008entropy} used an entropy-based multi-thresholding scheme to segment tumors. Gul et al.~(2012)\cite{li2012new} proposed a unified level set framework for semi-automatic liver tumor segmentation on contrast-enhanced CT. Other methods include texture-based deformable surfaces\cite{vorontsov2014metastatic}, Di fuzzy C-means with adaptive thresholding\cite{das2016kernelized}, graph-cut-based approaches\cite{liao2019automatic,wu20173d}, and multi-kernel clustering strategies\cite{krishan2019effective}.

With the development of machine learning, pixel-wise classification became a popular paradigm. Image patches centered at each pixel are described by handcrafted features (intensity statistics, texture, wavelets, etc.) and then classified by models such as support vector machines or random forests\cite{li2007machine,zhou2008semi,conze2017scale,huang2014random}. In addition, some works integrate learning with graph models, for example, by coupling tree-metric learning with graph cuts\cite{fang2011segmentation}, or by employing probabilistic graphical models and manifold learning for tumor segmentation\cite{hame2012semi,kadoury2015metastatic}. However, these approaches often require carefully designed features and are limited in capturing high-level semantic context.

\subsection{Deep Learning for Medical Image Segmentation}
Fully convolutional networks (FCNs)\cite{shelhamer2017fully} marked a turning point by enabling end-to-end dense prediction. Based on FCNs, Pinheiro et al.\cite{pinheiro2016learning} introduced refinement modules to improve object boundaries, and He et al.\cite{he2020mask} proposed Mask R-CNN, a widely used framework for instance segmentation. For medical images, U-Net\cite{ronneberger2015u} and its 3D variants\cite{cicek20163d,milletari2016v,korez2016model} have become standard baselines due to their encoder--decoder architecture with skip connections, which allows combining high-resolution spatial details with deep semantic features.

In the context of liver tumor segmentation, Vorontsov et al.\cite{vorontsov2018liver} adopted two cascaded FCNs for joint liver and lesion segmentation, while Pang et al.\cite{pang2019modified} improved the cascade by introducing dilated convolutions. Rezaei et al.\cite{rezaei2018instance} first detected tumor regions with an object detection CNN and then applied FCN within detected boxes. Li et al.\cite{li2018hdenseunet} proposed H-DenseUNet, which integrates dense connections into U-Net to enhance feature reuse and achieved strong performance on the LiTS challenge.

For multi-slice CT or MRI, the stack of slices can be considered as a short temporal sequence, motivating the use of recurrent neural networks (RNNs). Xu et al.\cite{xu2019lstm} exploited LSTM-enhanced U-Nets for brain tumor segmentation, and Liang et al.\cite{liang2019multi} used RNNs for liver tumor detection. Generative adversarial networks (GANs) have also been studied for medical segmentation\cite{grillo2020vox2vox,chen2019adversarial,yang2020segmentation,rezaei2020recurrent}, showing improvements in boundary realism and robustness.

\subsection{Radiomics and Deep Radiomics}
Radiomics\cite{aerts2014decoding,lambin2017radiomics} quantitatively extracts a large number of imaging features, such as texture, shape, and wavelet descriptors, from regions of interest, and builds models to predict tumor phenotype, treatment response, or prognosis. With the growth of imaging data and computational power, radiomics has become an important bridge between imaging and personalized medicine.

More recently, deep learning has been incorporated into radiomics workflows. CNNs can serve either as automatic feature extractors or as complementary feature sources to handcrafted descriptors\cite{afshar2019handcrafted,parekh2019radiomic}. Yang et al.\cite{yang2019probabilistic} further proposed probabilistic radiomics with multi-task learning to better model uncertainty and inter-observer variability.

Existing evidence suggests that when tumor contours are already reasonably accurate, handcrafted radiomics features still have advantages in describing shape and heterogeneity in a model-agnostic manner. In this work, we therefore use CNNs for coarse segmentation and boundary refinement, while relying on radiomics for robust false-positive suppression.

\section{Segmentation Method}

\subsection{Segmentation Model}
We propose a cascaded 2.5D U-Net architecture that jointly performs liver and liver-tumor segmentation in a single end-to-end trainable pipeline. Let $I \in \mathbb{R}^{H \times W \times Z}$ denote a CT volume with $Z$ axial slices. For each target slice index $z$, a three-slice stack
\[
X_z = \big[I_{z-1}, I_{z}, I_{z+1}\big] \in \mathbb{R}^{3 \times H \times W}
\]
is constructed as input, which allows the model to exploit limited through-plane context while remaining computationally efficient.

The framework comprises two highly modified U-Net branches sharing the same backbone design but serving different purposes. The first branch (liver network) takes $X_z$ as input and outputs a soft liver probability map $p^{\text{liv}}_z \in [0,1]^{H \times W}$. The second branch (tumor network) receives the identical three-slice stack and concatenates $p^{\text{liv}}_z$ as an additional fourth channel, forming
\[
\tilde{X}_z = \big[X_z,\, p^{\text{liv}}_z\big] \in \mathbb{R}^{4 \times H \times W}.
\]
This cascaded conditioning provides a strong anatomical prior, dramatically reducing search space for small and low-contrast tumors that would otherwise be overwhelmed by the dominant liver parenchyma.

The encoder of both branches follows the DenseNet-169 design \cite{huang2017densely}, consisting of four dense blocks (6, 12, 32, 24 dense layers respectively) interleaved with three transition-down layers. To keep the total model size manageable for joint training ($\sim$12M parameters per branch), we set the growth rate to 32 and apply a compression factor of 0.5 in transition layers. Dense connectivity encourages aggressive feature reuse, alleviates vanishing-gradient problems in very deep networks, and enables rich multi-scale feature representation with relatively few parameters---properties that are especially valuable when training data are scarce.

The decoder is symmetric and contains four upsampling stages. Starting from the $7 \times 7$ bottleneck, feature maps are progressively restored to $112 \times 112$ and finally to the original $224 \times 224$ resolution. Instead of conventional transposed convolution, which is prone to checkerboard artifacts, we adopt sub-pixel convolution (pixel shuffling) following Isola et al.\ \cite{isola2017image} and Shi et al.\ \cite{shi2016real}. This operation periodically rearranges $C \times H \times W$ low-resolution feature maps into a single $(C / r^{2}) \times (rH) \times (rW)$ high-resolution map (with upscale factor $r = 2$), yielding smoother boundaries and better preservation of high-frequency details critical for delineating irregular tumor margins.

Standard skip connections suffer from a semantic gap between high-resolution but low-semantic encoder features and low-resolution but high-semantic decoder features. To mitigate this, we introduce dense skip connections inspired by UNet++. Each encoder level not only forwards features to its corresponding decoder level but also generates intermediately upsampled versions that are fused (via concatenation followed by $3 \times 3$ convolution) into multiple deeper decoder stages. This nested horizontal and vertical dense connectivity greatly enhances information flow, strengthens semantic consistency across scales, and stabilises optimisation of the cascaded system.

To explicitly address the challenge of small and weakly contrasted tumors being suppressed in deep layers, we incorporate multi-scale attention gates (AG) \cite{oktay2018attention} into every skip pathway. The original 3D formulation is naturally adapted to our 2.5D setting by using $1 \times 1$ convolutions and bilinear resampling. For each skip connection, coarse-scale information from the decoder is used as a gating signal to compute spatial attention coefficients $\alpha \in [0,1]$, which are multiplied element-wise with the encoder features before concatenation. Concretely, for encoder feature map $x$ and gating feature map $g$, the attention coefficients are given by
\[
\alpha = \sigma\big( \psi^{\top} \big( \delta(W_x * x + W_g * g + b) \big) \big),
\]
where $W_x$ and $W_g$ are learnable $1 \times 1$ convolution kernels, $b$ is a bias term, $\delta(\cdot)$ denotes ReLU, and $\sigma(\cdot)$ is the Sigmoid function. Attention gates are applied at four different resolutions ($56 \times 56$, $28 \times 28$, $14 \times 14$, $7 \times 7$), enabling the network to dynamically focus on relevant lesion regions regardless of their size, shape, or contrast while suppressing irrelevant background parenchyma. This mechanism has proven particularly effective for detecting sub-centimetre metastases and diffuse infiltrative tumors.

\subsection{Loss Function}
The cascaded U-Net is trained end-to-end using a weighted combination of Dice loss and binary cross-entropy (BCE) for both liver and tumor branches. Let $p^{(c)}_i \in [0,1]$ denote the predicted probability at voxel $i$ for class $c \in \{\text{liver}, \text{tumor}\}$ and $t^{(c)}_i \in \{0,1\}$ the corresponding ground-truth label. The class-specific soft Dice loss is defined as
\[
\mathcal{L}^{(c)}_{\text{Dice}} = 1 - \frac{2 \sum_{i} p^{(c)}_i t^{(c)}_i + \epsilon}{\sum_{i} (p^{(c)}_i)^2 + \sum_{i} (t^{(c)}_i)^2 + \epsilon},
\]
where $\epsilon$ is a small constant for numerical stability. The BCE loss is given by
\[
\mathcal{L}^{(c)}_{\text{BCE}} = - \frac{1}{|\Omega|} \sum_{i \in \Omega} \big[ t^{(c)}_i \log p^{(c)}_i + (1 - t^{(c)}_i) \log (1 - p^{(c)}_i) \big],
\]
where $\Omega$ denotes the set of voxels within the field of view.

The total segmentation loss aggregates the contributions from both tasks:
\[
\mathcal{L}_{\text{seg}} = \sum_{c \in \{\text{liver}, \text{tumor}\}} w_c \Big( \lambda_{\text{Dice}} \, \mathcal{L}^{(c)}_{\text{Dice}} + \lambda_{\text{BCE}} \, \mathcal{L}^{(c)}_{\text{BCE}} \Big),
\]
where $w_c$ are task weights. In all experiments, we set $\lambda_{\text{Dice}} = 0.7$, $\lambda_{\text{BCE}} = 0.3$, $w_{\text{liver}} = 1$, and $w_{\text{tumor}} = 2$ to alleviate the strong class imbalance between liver parenchyma and tumor tissue.

\subsection{Post-processing of Segmentation}

\subsubsection{Morphological Smoothing and Temporal Consistency}

The cascaded network generates independent per-pixel probability maps for the liver and tumors, with values normalised to $[0, 1]$. For implementation convenience, these volumes are first linearly scaled to the 8-bit range $[0, 255]$. Rather than applying a fixed threshold, we employ Otsu's adaptive thresholding algorithm \cite{otsu1979threshold} independently for the liver and tumor channels of each volume.

Let $h(k)$ denote the histogram of gray levels $k \in \{0, \dots, 255\}$, normalised such that $\sum_{k} h(k) = 1$. For a candidate threshold $\tau$, define
\[
\omega_0(\tau) = \sum_{k=0}^{\tau} h(k), \quad
\omega_1(\tau) = \sum_{k=\tau+1}^{255} h(k),
\]
and the corresponding class means
\[
\mu_0(\tau) = \frac{1}{\omega_0(\tau)} \sum_{k=0}^{\tau} k \, h(k), \quad
\mu_1(\tau) = \frac{1}{\omega_1(\tau)} \sum_{k=\tau+1}^{255} k \, h(k).
\]
Otsu's method selects the threshold
\[
\tau^{\star} = \arg\max_{\tau} \, \omega_0(\tau)\,\omega_1(\tau)\,\big[\mu_0(\tau) - \mu_1(\tau)\big]^2,
\]
which maximises the between-class variance and yields a case-specific threshold that effectively handles the substantial intensity and contrast variations observed across different scanners, reconstruction kernels, and contrast phases in real-world abdominal CT datasets.

After thresholding, let $B^{(c)} \in \{0,1\}^{H \times W \times Z}$ denote the resulting binary mask for class $c \in \{\text{liver}, \text{tumor}\}$. A lightweight morphological cleaning step is then performed using a $3\times3$ square structuring element $S$. We apply a morphological closing operation
\[
\hat{B}^{(c)} = \big(B^{(c)} \ominus S\big) \oplus S,
\]
where $\ominus$ and $\oplus$ denote binary erosion and dilation, respectively. This process efficiently removes isolated false-positive voxels and small speckles that typically arise from minor probability fluctuations near class boundaries, while simultaneously smoothing minor contour irregularities. Quantitative analysis on the LiTS and our in-house datasets shows that this step eliminates over 92\% of connected components smaller than 8 voxels without compromising the global Dice coefficient.

Despite its effectiveness against noise, aggressive erosion can inadvertently suppress genuine small metastases (typically $<6$ mm in diameter) or fracture thin, vessel-like tumor extensions. To counteract these undesirable side effects without reintroducing noise, we exploit the strong anatomical continuity along the $z$-axis that characterises portal-venous phase abdominal CT volumes (slice thickness 1--3 mm, minimal respiratory motion).

We introduce a simple yet highly effective three-slice consistency refinement rule specifically targeted at the tumor mask. For each voxel at position $(x, y, z)$ in the tumor volume, let $y_{x,y,z} \in \{0,1\}$ denote the current binary label after morphological cleaning, and let $p_{x,y,z} \in [0,1]$ denote the original soft probability before thresholding. The refined label $\tilde{y}_{x,y,z}$ is defined as
\[
\tilde{y}_{x,y,z} =
\begin{cases}
1, & \text{if } y_{x,y,z} = 1, \\[4pt]
1, & \text{if } y_{x,y,z} = 0,\ y_{x,y,z-1} = 1,\ y_{x,y,z+1} = 1, \\[-2pt]
   & \qquad \text{and } \dfrac{p_{x,y,z-1} + p_{x,y,z+1}}{2} > 0.6, \\[6pt]
0, & \text{otherwise.}
\end{cases}
\]

Intuitively, voxels that are labelled as tumor in both the preceding slice $z-1$ and the following slice $z+1$, but are absent (background) in the current slice $z$, are restored if the average raw probability of the corresponding location in the two neighbouring slices exceeds 0.6. Conversely, isolated tumor voxels that lack foreground support from both adjacent slices are suppressed. This bidirectional rule functions as a robust median-like filter along the cranio-caudal direction and requires only two lightweight passes over the entire volume.

\subsubsection{Removing False Positives with Radiomics}

\paragraph{Sampling Strategy}
To train a robust classifier for false-positive removal, both positive and negative 3D tumor candidates are required. Positive samples are obtained from expert-annotated tumor volumes. Negative samples are drawn from normal liver parenchyma, liver boundary regions in contact with other organs, and regions containing only a small fraction of lesion voxels.

The dataset contains CT scans of 300 patients. We adopt 5-fold cross-validation, using 80\% of the cases (240 patients) for training in each fold. This yields approximately 1{,}000--1{,}300 annotated tumor regions per training fold. To maintain class balance, 1{,}000 negative regions are sampled as follows:

\begin{enumerate}
\item Load a patient's CT and liver/tumor masks.
\item Initialize sampling radius $r=2$ pixels with step size $s=2$ pixels.
\item Randomly select non-tumor seed voxels near the liver boundary.
\item Randomly select non-tumor seed voxels inside the liver parenchyma.
\item For each seed, extract a 3D patch of radius $r$.
\item Discard the patch if more than 20\% of voxels lie outside the liver or more than 10\% belong to tumor. Steps (3)--(5) are repeated up to 100 times for each radius.
\item Increase radius $r \leftarrow r+s$ and repeat until $r=48$.
\item Proceed to the next CT volume until all scans are processed.
\end{enumerate}

The radius thus ranges from 2 to 48 pixels, resulting in 24 scales. To approximate a uniform size distribution, each scale contributes roughly $1{,}000/24 \approx 42$ negative samples. For each scale, 40\% of patches are selected from intra-hepatic regions and 60\% from boundary regions, reflecting the higher propensity for false positives around organ interfaces.

\paragraph{Radiomic Feature Extraction}
Within each candidate region $\Omega_k$, we compute eight groups of radiomic features, which together form a 728-dimensional feature vector $\phi(\Omega_k) \in \mathbb{R}^{728}$:

\begin{enumerate}
\item \textbf{Histogram features}: 200-bin intensity histograms with bin width 4 are used to derive mean, standard deviation, skewness, kurtosis, energy, entropy, and other first-order descriptors.

\item \textbf{Gradient features}: 3D gradients are computed via Gaussian convolution with standard deviation $\sigma$ set to 1.5 times the voxel spacing. The mean and standard deviation of gradient magnitude within the region are used.

\item \textbf{Run-length (RL) texture features}: Eleven RL features are derived from a 128-bin RL matrix, which encodes the number of runs with gray level $i$ and length $j$. Fine textures exhibit more short runs, while coarse textures have more long runs.

\item \textbf{Gray-level co-occurrence matrix (GLCM) features}: Twenty-two GLCM features are extracted from 128-bin matrices to characterize spatial gray-level dependence.

\item \textbf{Shape features}: Tumor shape is quantified by descriptors such as volume, surface area, sphericity, and compactness.

\item \textbf{Second-order moment invariants}: Three central moment invariants $J_1$, $J_2$, and $J_3$ capture the global spatial distribution of intensities.

\item \textbf{Boundary features}: To emphasize heterogeneity at tumor margins, the above features (except shape) are recomputed within a narrow band around the boundary.

\item \textbf{3D wavelet features}: 3D wavelet decomposition is performed along three axes, with low- and high-frequency components in each direction, resulting in eight sub-bands. For each sub-band, non-shape features are recalculated.
\end{enumerate}

In total, 728 features are obtained for each candidate region.

\paragraph{Feature Selection and Classification}
All features are standardized to zero mean and unit variance. Features with near-zero variance are discarded. Pairwise Pearson and distance correlation coefficients are then computed to remove strongly correlated features, reducing redundancy.

To further refine the feature set, we apply six selection strategies: recursive feature elimination (RFE), Lasso regression, random forest importance, XGBoost importance, gradient boosting decision trees (GBDT), and Relief. For each method, the top 30 features are retained. The intersection of the six selected sets yields 20 stable, highly informative features, denoted by $\tilde{\phi}(\Omega_k) \in \mathbb{R}^{20}$.

Given their strong performance in radiomics literature, we adopt a random forest classifier as the final model. Let $f_{\theta}$ denote the trained random forest with parameters $\theta$. For each candidate region, the probability of being a true tumor is estimated as
\[
q_k = f_{\theta}\big(\tilde{\phi}(\Omega_k)\big) \in [0,1].
\]
During inference, candidates with $q_k$ below a fixed threshold $\tau_{\text{rf}}$ are suppressed. As shown in Section~\ref{sec:experiments}, the random forest achieves the best balance between accuracy and robustness among the tested classifiers.

\subsubsection{Edge Refinement with 3D CNN}
Even after radiomics-based filtering, the segmentation masks may contain slightly over-segmented regions or jagged contours. To obtain smoother and more anatomically plausible boundaries, we perform a final voxel-level refinement.

For each tumor, we construct a narrow band $\Gamma$ around the current boundary by applying a signed distance transform $d(\cdot)$ to the binary tumor mask and selecting voxels with $|d(\mathbf{v})| \leq d_{\max}$ (here $d_{\max}=6$ pixels). Voxels inside the tumor (negative distance) are labeled as positive, and voxels within a distance of six pixels outside the boundary (positive distance) are labeled as negative. For each labeled voxel $\mathbf{v}$, a 3D image block $P(\mathbf{v}) \in \mathbb{R}^{s \times s \times s}$ centered at $\mathbf{v}$ is extracted as input, where $s$ denotes the patch size. Regions far away from the tumor are not sampled, as they are not relevant for local boundary correction.

A compact 3D CNN inspired by AlexNet is designed for this task. CT images are single-channel, and experiments show no clear benefit from replicating or transforming them into three channels. Therefore, the network takes a one-channel 3D patch as input. Patch sizes between $5 \times 5 \times 5$ and $21 \times 21 \times 21$ are evaluated; an $11 \times 11 \times 11$ patch yields the best trade-off between context and trainability.

All convolution layers use $3 \times 3 \times 3$ kernels with stride 1, followed by $2 \times 2 \times 2$ max-pooling. The network consists of two 64-channel conv layers, two 128-channel conv layers, and one 256-channel conv layer, followed by fully connected layers and a final Sigmoid output. Compared with very deep models such as VGG, Inception, or ResNet, this architecture is shallower and faster to train, yet sufficient to capture local boundary patterns.

During inference, the classifier is applied to voxels within the boundary band $\Gamma$, and tumor labels are updated accordingly, leading to refined contours.

\section{Experiments and Discussion}
\label{sec:experiments}
All experiments are conducted on an NVIDIA Tesla P100-PCIE-12GB GPU under Ubuntu 16.04 using Python 3.6.5. The main libraries are TensorFlow 1.12.0, Keras 2.2.5, and cuDNN 7.1.4. Unless otherwise stated, all reported numbers are averaged over five folds with the corresponding standard deviation.

\subsection{Datasets and Preprocessing}
The in-house liver cancer CT dataset is collected from several collaborating hospitals and includes 300 patients with pathologically or clinically confirmed primary or metastatic liver tumors. For each case, a portal-venous phase abdominal CT volume is available. The in-plane resolution ranges from 0.6 to 0.9 mm, and the slice thickness ranges from 1.0 to 3.0 mm, depending on the scanner and acquisition protocol.

All volumes are resampled to an isotropic in-plane spacing of 1.0 mm using third-order B-spline interpolation and then cropped or padded to $224 \times 224$ pixels around the liver region based on a coarse body mask. Intensities are clipped to the range [$-100$, 400] Hounsfield units and linearly rescaled to [0, 255]. The rescaled images are stored as 8-bit grayscale slices to reduce storage requirements and accelerate training.

Liver and tumor masks are manually delineated slice by slice by radiologists with more than 10 years of experience in abdominal imaging. For cases with ambiguous tumor boundaries, annotations are obtained by consensus after consultation among multiple experts. To minimise potential bias from annotation noise, the consensus masks are used as the ground truth in all experiments.

We adopt five-fold cross-validation on the 300-patient cohort. In each fold, 240 cases are used for training, 20 for validation (for model selection and early stopping), and 40 for testing. The validation set is randomly sampled from the training pool at the beginning of each fold and kept fixed throughout the training process. For the segmentation networks, 25 representative slices per volume are selected for training based on liver coverage and presence of tumor tissue, resulting in 6{,}000 slices per fold. To isolate the effect of the proposed architectural components, no data augmentation (such as rotation, scaling, or elastic deformation) is used in this study.

For external evaluation, we further test the trained models on the LiTS benchmark\cite{rezaei2018instance}. Following standard practice, LiTS volumes are resampled and intensity-normalized using the same protocol as the in-house data. Ground-truth liver and tumor masks from LiTS are used without modification.

\subsection{Training Protocol and Implementation Details}
The cascaded U-Net is trained end-to-end using a weighted combination of Dice loss and binary cross-entropy for both liver and tumor branches. Specifically, the total loss is defined as
\[
\mathcal{L} = \lambda_{\text{Dice}} \, \mathcal{L}_{\text{Dice}} + \lambda_{\text{BCE}} \, \mathcal{L}_{\text{BCE}},
\]
where $\lambda_{\text{Dice}} = 0.7$ and $\lambda_{\text{BCE}} = 0.3$ in all experiments. To alleviate class imbalance between foreground (tumor) and background, the tumor loss is scaled by a factor of 2 relative to the liver loss.

We use Adam with an initial learning rate of $3 \times 10^{-4}$ and an exponential decay of $5 \times 10^{-5}$. The batch size is set to 24 for the cascaded segmentation networks. Batch normalisation and ReLU activation are used after each convolution layer except the final prediction layer, which uses Sigmoid activation. A model checkpoint callback saves the weights with the lowest validation loss. The maximum number of epochs is 160, and early stopping is triggered if the validation loss does not improve for 12 consecutive epochs.

For the radiomics classifier, feature standardisation parameters (mean and variance) are estimated on the training set of each fold and applied to the corresponding validation and test sets. The random forest classifier is trained using 350 trees with a maximum depth of 16 and a minimum of 10 samples per leaf. Class weights are balanced to account for potential minor residual class imbalance after sampling.

The 3D CNN for boundary refinement is trained using balanced positive and negative samples (50{,}000 each per fold). We first train with Adam (learning rate $1 \times 10^{-3}$) for 30 epochs to obtain a good initial solution and then switch to stochastic gradient descent (learning rate $1 \times 10^{-4}$, momentum 0.9) for fine-tuning. The batch size is 32 for patch sizes up to $15 \times 15 \times 15$ and 16 for larger patches. Early stopping with a patience of 10 epochs is applied based on validation loss.

\subsection{False-Positive Removal: Classifier Comparison}
Table~\ref{t1} summarises the performance of different classifiers for distinguishing true tumors from false positives using the selected radiomic features. Hyperparameters are tuned via grid search within each model family. All results are averaged over the five cross-validation folds.

\begin{table}[htbp]
\centering
\caption{Comparison of different classification models and their cross-validated accuracy}
\begin{tabular}{lcc}
\toprule
\textbf{Classification Model} & \textbf{Key Hyperparameters} & \textbf{Cross-Validation Accuracy} \\
\midrule
Random Forest          & n\_trees = 350                  & $0.95 \pm 0.03$ \\
SVM (RBF kernel)       & C = 0.2, gamma = 0.4            & $0.92 \pm 0.04$ \\
SVM (linear kernel)    & C = 1.0                         & $0.88 \pm 0.04$ \\
Logistic Regression    & penalty = L2, C = 0.001         & $0.81 \pm 0.07$ \\
\bottomrule
\label{t1}
\end{tabular}
\end{table}

The random forest achieves the highest accuracy and the most stable performance across folds. Receiver operating characteristic (ROC) analysis further shows that the random forest yields an AUC close to 0.99, clearly outperforming the other classifiers. In addition, we observe that the random forest is less sensitive to the choice of the feature subset size than sparse linear models such as Lasso, which supports its use as the default classifier in the final pipeline.

\subsection{3D CNN for Edge Refinement}
To evaluate the effectiveness of the 3D CNN architecture for boundary refinement, we compare the adapted AlexNet design with deeper CNNs under the same training protocol. The input patch size is fixed to $15 \times 15 \times 15$ in this experiment, and the number of positive and negative samples is balanced at 50{,}000 each. Results are reported in Table~\ref{t2}.

\begin{table}[!htbp]
\centering
\caption{Performance comparison of different CNN models (patch size $15 \times 15 \times 15$)}
\begin{tabular}{lcc}
\toprule
\textbf{CNN Model} & \textbf{Accuracy} & \textbf{Training Speed} \\
\midrule
AlexNet (3D adapted) & 0.97              & Fast                    \\
ResVNet (3D adapted) & 0.92              & Medium                  \\
MobileNet (3D)       & 0.92              & Slow                    \\
ResNet50 (3D)        & 0.94              & Slow                    \\
\bottomrule
\end{tabular}
\addvspace{8pt}
\label{t2}
\end{table}

Although deeper models achieve competitive accuracy, their training time and memory consumption are significantly higher. The adapted AlexNet offers a good balance between performance and efficiency and is thus selected for edge refinement in the proposed framework.

We further investigate the influence of patch size on classification performance. Table~\ref{t3} lists the results for different cubic patch sizes using the adapted AlexNet. The best performance is obtained with an $11 \times 11 \times 11$ patch, confirming that this size provides sufficient local context without introducing excessive complexity.

\begin{table}[!htbp]
\centering
\caption{Performance comparison of different image block sizes for boundary classification}
\begin{tabular}{cc}
\toprule
\textbf{Block Size} & \textbf{Accuracy} \\
\midrule
$9 \times 9 \times 9$    & 0.96 \\
$11 \times 11 \times 11$ & \textbf{0.98} \\
$13 \times 13 \times 13$ & 0.97 \\
$15 \times 15 \times 15$ & 0.96 \\
$23 \times 23 \times 23$ & Failed to converge \\
\bottomrule
\end{tabular}
\addvspace{8pt}
\label{t3}
\end{table}

When integrated into the full segmentation pipeline, the 3D CNN refinement yields an average improvement of about 0.6 percentage points in tumor Dice on the in-house dataset and 0.4 percentage points on LiTS, with more obvious gains in cases with irregular or highly heterogeneous tumor boundaries.

\subsection{Segmentation Metrics and Comparative Evaluation}
Segmentation performance is evaluated using sensitivity, positive predictive value (PPV), and Dice coefficient (DSC):
\[
R_{\text{sensitivity}} = \frac{| P \cap T |}{| T |}, \quad
R_{\text{PPV}} = \frac{| P \cap T |}{| P |},
\]
\[
R_{DC} = \frac{2 | P \cap T |}{| P | + | T |},
\]
where $P$ and $T$ denote the predicted and ground-truth tumor voxels, respectively. All metrics are computed at the volume level and then averaged across all test cases in each fold.

We compare the proposed method with three representative approaches: Pang et al.~(2019)\cite{pang2019modified}, Vorontsov et al.~(2018)\cite{vorontsov2018liver}, and Li et al.~(2018)\cite{li2018hdenseunet}. For a fair comparison, all methods are reimplemented or adapted under the same preprocessing, training, and evaluation protocols where possible. The results on both the in-house hospital dataset and the LiTS benchmark\cite{rezaei2018instance} are reported in Table~\ref{t4}.

\begin{table}[!htbp]
\centering
\caption{Comparison of segmentation performance with state-of-the-art methods}
\begin{tabular}{l *{8}{c}}
\toprule
\textbf{Metric} & \multicolumn{2}{c}{\textbf{Proposed}} & \multicolumn{2}{c}{\textbf{Pang et al. (2019)\cite{pang2019modified}}} & \multicolumn{2}{c}{\textbf{Vorontsov et al. (2018)\cite{vorontsov2018liver}}} & \multicolumn{2}{c}{\textbf{Li et al. (2018)\cite{li2018hdenseunet}}} \\
\cmidrule(lr){2-3} \cmidrule(lr){4-5} \cmidrule(lr){6-7} \cmidrule(lr){8-9}
& Hospital & LiTS & Hospital & LiTS & Hospital & LiTS & Hospital & LiTS \\
\midrule
Sensitivity   & \textbf{$0.89 \pm 0.02$} & \textbf{0.86} & $0.81 \pm 0.03$ & 0.80 & $0.83 \pm 0.04$ & 0.80 & $0.86 \pm 0.02$ & 0.86 \\
PPV           & \textbf{$0.92 \pm 0.02$} & \textbf{0.90} & $0.85 \pm 0.03$ & 0.84 & $0.86 \pm 0.05$ & 0.83 & $0.90 \pm 0.02$ & 0.90 \\
DSC           & \textbf{$0.88 \pm 0.03$} & \textbf{0.84} & $0.79 \pm 0.02$ & 0.79 & $0.81 \pm 0.04$ & 0.79 & $0.84 \pm 0.02$ & 0.83 \\
\bottomrule
\label{t4}
\end{tabular}
\end{table}

On the hospital dataset, the proposed method improves sensitivity by 0.03, PPV by 0.02, and Dice by 0.04 compared with the strong baseline H-DenseUNet\cite{li2018hdenseunet}. Gains over cascaded FCN-based approaches\cite{vorontsov2018liver,pang2019modified} are even larger, particularly in terms of Dice. Similar trends are observed on the LiTS dataset, indicating good generalization to unseen data and different acquisition protocols. A paired Wilcoxon signed-rank test between the proposed method and H-DenseUNet yields $p < 0.01$ for both Dice and sensitivity on the hospital dataset, suggesting that the improvements are statistically significant.

\subsection{Ablation Study}
To better understand the contribution of each component in the proposed pipeline, we conduct an ablation study on the in-house dataset. Starting from a baseline cascaded U-Net with standard skip connections, bilinear upsampling, and no attention, temporal refinement, radiomics, or 3D CNN, we progressively add the proposed modules. Table~\ref{t5} reports the resulting tumor Dice scores.

\begin{table}[!htbp]
\centering
\caption{Ablation study of major components on the hospital dataset (tumor DSC)}
\begin{tabular}{l c}
\toprule
\textbf{Configuration} & \textbf{DSC} \\
\midrule
Baseline cascaded U-Net                                             & $0.82 \pm 0.03$ \\
+ Dense encoder/decoder and sub-pixel upsampling                    & $0.84 \pm 0.03$ \\
+ Attention gates on skip connections                               & $0.86 \pm 0.03$ \\
+ Temporal consistency refinement                                   & $0.87 \pm 0.03$ \\
+ Radiomics-based false-positive removal                            & $0.87 \pm 0.02$ \\
+ 3D CNN boundary refinement (full proposed pipeline)               & $\mathbf{0.88 \pm 0.03}$ \\
\bottomrule
\label{t5}
\end{tabular}
\end{table}

The results reveal several important trends:

\begin{itemize}
\item Introducing dense encoder/decoder connections and sub-pixel upsampling yields a clear improvement of 0.02 in Dice, highlighting the importance of multi-scale feature reuse and high-quality resolution recovery.

\item Adding attention gates further improves Dice by 0.02. Visual inspection shows that this module is particularly beneficial for detecting small satellite lesions and suppressing irrelevant high-intensity structures near the liver boundary.

\item Temporal consistency refinement brings an additional 0.01 increase in Dice while recovering many small metastases that were partially removed by morphological operations.

\item Radiomics-based false-positive removal substantially reduces spurious detections in highly heterogeneous regions. Its impact on global Dice is modest (approximately 0.0--0.01) because the rejected regions typically occupy a small volume, but it significantly improves lesion-level precision (not shown in the table).

\item 3D CNN boundary refinement provides the final 0.01 gain in Dice and leads to visually smoother and more anatomically plausible contours.
\end{itemize}

Overall, the ablation study demonstrates that each component contributes to the final performance, with dense connections, attention gates, and temporal consistency being the most influential for voxel-wise accuracy.

\subsection{Performance on Tumors of Different Sizes}
Accurate segmentation of small tumors is clinically important but technically challenging. To evaluate the size-dependent performance, we stratify tumors in the hospital dataset into three groups based on their equivalent spherical diameter: small ($<$10 mm), medium (10--30 mm), and large ($>$30 mm). Table~\ref{t6} compares the Dice scores of the proposed method and H-DenseUNet\cite{li2018hdenseunet} for each group.

\begin{table}[!htbp]
\centering
\caption{Size-stratified tumor Dice on the hospital dataset}
\begin{tabular}{lccc}
\toprule
\textbf{Method} & \textbf{Small ($<$10 mm)} & \textbf{Medium (10--30 mm)} & \textbf{Large ($>$30 mm)} \\
\midrule
H-DenseUNet\cite{li2018hdenseunet} & $0.68 \pm 0.08$ & $0.84 \pm 0.03$ & $0.91 \pm 0.02$ \\
Proposed                            & $\mathbf{0.74 \pm 0.07}$ & $\mathbf{0.87 \pm 0.03}$ & $\mathbf{0.92 \pm 0.02}$ \\
\bottomrule
\label{t6}
\end{tabular}
\end{table}

The proposed method achieves the largest relative improvement for small lesions, with a 0.06 increase in Dice compared with H-DenseUNet. This is mainly attributed to the attention gates and dense skip connections, which enhance the representation of small-scale features, as well as the temporal refinement that helps restore small lesions across slices. For medium and large tumors, the gains are smaller but still consistent, indicating that the proposed enhancements do not compromise performance on easier cases.

\subsection{Robustness Analysis and Qualitative Observations}
To assess robustness to imaging variations, we perform several additional tests. First, we synthetically perturb test volumes with Gaussian noise (standard deviation up to 10 HU) and random global intensity scaling (within $\pm$10\%) to mimic differences in scanner noise and contrast injection. The overall Dice score of the proposed method on the hospital dataset decreases by less than 1.0 percentage point under the strongest perturbation, suggesting good robustness to moderate acquisition variability.

Second, we evaluate the sensitivity of the radiomics classifier to the number of selected features. When varying the final feature dimensionality between 12 and 30, the classification accuracy remains above 0.93, and the impact on segmentation Dice is within 0.3 percentage points. This indicates that the feature selection process is relatively stable and that the random forest can tolerate minor variations in the feature set.

Visual inspection of representative cases confirms the quantitative findings. All methods can segment large, well-contrasted lesions, but the proposed approach better preserves small satellite lesions and reduces over-segmentation in heterogeneous areas. The attention-enhanced dense U-Net captures subtle hypodense regions that are often missed by purely FCN-based methods, while radiomics-based filtering and 3D CNN refinement effectively remove spurious detections along vessel walls and organ interfaces and sharpen tumor boundaries.

From a clinical perspective, the proposed pipeline tends to slightly underestimate tumor volume in highly infiltrative cases but still captures the overall extent better than the comparison methods. In focal, well-defined tumors, the refined contours closely follow the expert annotations and are suitable for direct use in 3D visualisation, preoperative planning, and quantitative volumetric analysis.

\section{Conclusion}
This paper proposes a hybrid liver and liver-tumor segmentation framework that combines an enhanced cascaded U-Net, handcrafted radiomics, and 3D voxel-block CNN refinement. The segmentation network introduces multi-scale dense skip connections, sub-pixel convolution, and attention gates to better capture small, low-contrast lesions. Radiomic features are used to train a random forest classifier that effectively suppresses false-positive regions, and a compact 3D CNN refines tumor boundaries in a narrow band around the contour.

Experiments on 300 clinically annotated CT cases and the LiTS dataset demonstrate that the proposed method consistently outperforms several state-of-the-art approaches in terms of sensitivity, PPV, and Dice coefficient. The method has been integrated into a medical imaging research platform and can support 3D visualization, quantitative analysis, and preoperative planning.

Nevertheless, the current system requires training and maintaining three separate models, which complicates deployment and does not provide end-to-end optimization. In future work, we plan to explore GAN-based architectures\cite{yang2020segmentation,rezaei2020recurrent} to further enhance generalization and boundary realism and investigate how to embed radiomics-like feature reasoning into a unified end-to-end trainable network. Such integration may simplify the pipeline, improve computational efficiency, and facilitate broader clinical adoption.

\bibliographystyle{unsrt}  
\bibliography{accu_ref}
\end{document}